# Emissivity measurements with an Atomic Force Microscope

**P J van Zwol, L Ranno, J Chevrier**
*Institut Néel, CNRS and Universite Joseph Fourier Grenoble, BP 166 38042 Grenoble Cedex 9, France*

We show that functionalized micromechanical bilayer levers can be used as sensitive probes to accurately measure radiative heat flux in vacuum between two materials at the micro scale. By means of calibration to one material these measurements can be made quantitative for radiative heat flux or for either temperature or material emissivity. We discuss issues and opportunities for our method and provide ample technical details regarding its implementation and demonstrate good correspondence with the Stefan Boltzman law. We use this system to probe the phase transition of $VO_2$ and find that radiative heat transfer in farfield between $VO_2$ and glass can be reversibly modulated by a factor of 5.

## 1. Introduction

Precise knowledge of infrared absorption, thermal transfer and emissivity is important in solar and space related fields [1]. A critical factor for space applications is the reduction of size of thermal sensors. Low cost and small size thermal sensors also benefit appliances in domestic medical and military fields. On the other hand, as transistor and current densities are ever increasing, local determination of temperature at the micro or nano scale is of obvious interest in the microelectronics industry [2]. Hence, cost savings, and the ability to probe the temperature at the micro scale both give a strong incentive for decreasing the size, and increasing the sensitivity and accuracy of heat sensors.

Imaging techniques such as Infrared Thermography and Thermo Reflectance Microscopy are popular techniques that can probe the temperature or emissivity of a sample. [3]. The former is based on blackbody radiation of samples the latter does not probe emissivity but is based on temperature dependence of the reflection coefficient in the visible range. The first is most suited for low reflectivity materials whereas the latter is more suited for higher reflectivity materials. [3]. Infrared Thermography is popular and often utilized but a disadvantage is that IR photon detectors require cryogenic cooling, which drives up the size and cost of the device substantially [4].

Atomic Force Microscope (AFM) levers are micromechanical devices that have the potential to replace the two above mentioned techniques as a small cost effective alternative. They were already used in the 1980s in the form of Scanning Thermal Microscopy, a form of Scanning Probe Microscopy (SPM) [2,5]. The first probes consisted of a tungsten tip with a thermocouple at the apex. In this technique the tip is heated and brought in near contact with the substrate, which results in cooling of the tip. A feedback on the distance between tip and sample that keeps the temperature on the tip constant, allows the measurement of surface topology. Instead of thermocouples also resistive probes are used [2]. Usually these devices work in contact, but recently it has been used to probe near field radiative heat transfer in vacuum as well [6].

Near field Scanning Optical Microscopes or NSOMs are another class of AFM devices that allow, besides topology mapping, the scrutinization of thermal properties. Usually these devices work with visible light and are used as a near field analogue of a Thermo Reflectance Microscope [3]. Recently a near field NSOM based device where the probe scatters nearfield IR emissions into a sensor positioned at farfield has been used to probe thermally excited surface polaritons at the micro scale. Thus scanning probe techniques that probe radiative heat transfer, whether they are a variant of SPM [6] or NSOM [7], are relatively recent phenomena and work hitherto foremost in the near field regime.

In parallel to the development of thermal SPM and NSOM techniques, another AFM approach was developed at the end of the last century. It relies on the high sensitivity of micro machined bilayer sensors [8]. Such sensors bend due to heat flux, and can serve as sensitive calorimeters [8]. Bimorphs have been used as a scanning thermal (contact) technique to probe the local temperature of a sample at the micro scale [9, 10], and are finding new applications in the form of MEMS devices that should one day replace costly cryogenically cooled CCDs in IR Thermography [11, 12]. Recently bilayer sensors with attached micro spheres were used as a probe in a conventional AFM to measure radiative heat transfer [13] in near field. Not long after, an improved setup was presented which included extensive techniques to stabilize the heat flux from the laser on the lever [14]. This improved setup has been used to perform precise measurements of radiative thermal transfer in near field.

In this work we show how this setup can be used in the farfield regime. We discuss its peculiarities and applicable range and use it to measure thermal transfer of a material that undergoes a phase transition. The extensive measures that reduce the effects of energy transfer of the optical detection scheme on the bilayer probe, enable quantitative measurements of radiative heat transfer in near and farfield, which to our knowledge are at the moment of writing, unique to our setup. By means of calibration to a known material the temperature or emissivity of a sample can be determined with controlled precision. We believe that our results have importance for the development of novel IR detectors [11,12] and IR sensitive microprobes, as optical readout of lever motion remains a popular frequently used non contact technique, even in the case of MEMS arrays of micro levers [11]. We further believe it can complement existing SPM and NSOM techniques for quantitative temperature, emissivity or heat transfer measurements.

## 2. Experimental setup

Our setup is a high vacuum (between $5 \cdot 10^{-8}$ and $10^{-7}$mbar) AFM where the lever is rotated by 90 degrees to minimize the effect of electrostatic and dispersion forces (fig. 1). The bilayer lever is sensitive to heat flux due to the different expansion coefficient of the materials [8]. For small deflections the lever motion is linearly proportional to heat flux. Typically commercially available micro levers are sensitive to heat fluxes of 100pW [8]. Here we use 320 micron Veeco silicon nitride levers that consist of 500nm silicon nitride and 60nm Gold/Chromium, to which a sphere (diameter 40µm, Duke Scientific 9040) is attached. The lever remains stationary and resides above an XYZ stage of Attocube motors, which support the heater and the sample.

Obviously the laser in an AFM that senses lever motion induces a significant energy flux on the lever. This flux is much larger (tunable in the micro to milli Watt range) than the thermal transfer between a microsphere and a plate (typically in the nano Watt range). However, only 4% laser light is absorbed, because the gold layer on the lever is a good reflector [13]. Here we diminish laser flux related spurious signals by use of a a fiber interferometric setup for which the laser flux is kept constant by using a closed loop system and a low noise RF-modulated constant power laser source. The used laser power is about 0.1mW as was measured by a calibrated photo-detector, and compared to the maximum laser output (6mW). Following the arguments in ref. [13] we infer that the heating of the free end of the lever by the laser must therefore be about three Kelvin. The fiber is mounted on another XYZ stage of Attocube motors where a feedback loop on the X-piezo keeps the distance between the fiber and the lever constant (fig. 1). In this way the lever motion is precisely measured, and the heat flux of the laser on the lever is kept constant.

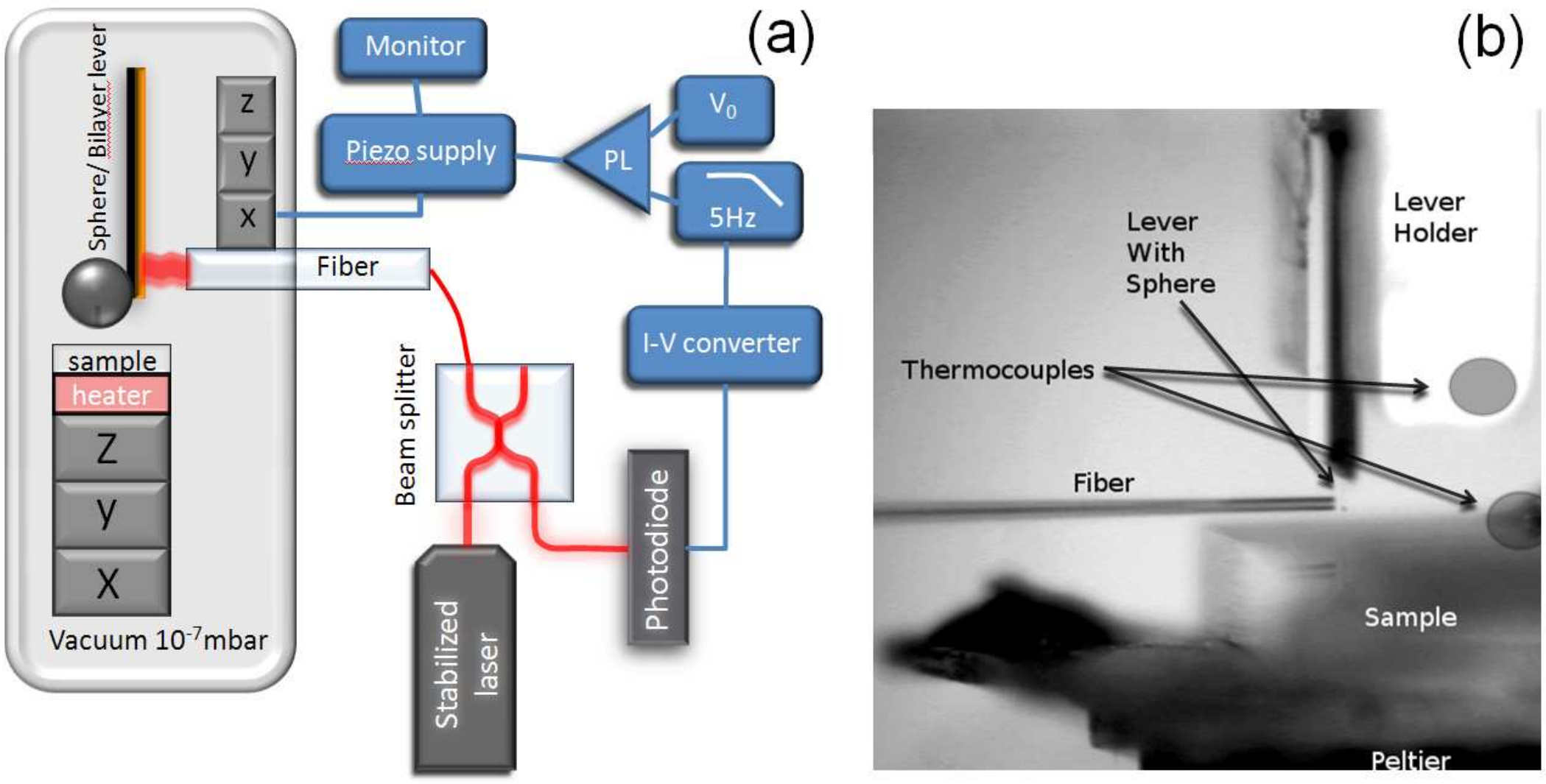

**Figure (1). (a)** Schematic of our setup with feedback loop. Both lever and sample are put on XYZ stages of Attocube motors (ANP100 series) in vacuum. A laser (Schäfter Kirchhoff 51nanoFI-660nm) is coupled to a 50-50 beam splitter (Schafter Kirchoff) from where a fiber goes into the vacuum chamber. The interference between the fiber and lever is sensed by a photodiode (Thorlabs DET100A). The signal of the photodiode is amplified by a Femto DLPCA 200 amplifier and goes into a Nanonis station with RC4 and SC4 real time and signal conditioning controllers, OC4 oscillation controller with integrated Phase Lock Loop, and HVA4 Piezo Supply that controls the X-piezo on which the fiber is mounted. **(b)** Photo of lever and sample as taken with a microscope. Different parts are indicated that are referred to in the text. The positions of the thermocouples are indicated by the circles.

A Peltier serves to heat the sample. The Peltier and sample have a size of $0.5cm^2$ typically. A thermocouple is mounted directly onto the sample with silver paste to measure its temperature. Another thermocouple is mounted on the lever holder (fig. 1b), since the Peltier heats not only the lever, but also the surroundings. All parts in our setup are largely metallic, including the motors. The Attocube motors and aluminum lever holder are mounted on an aluminum support. This ensures an optimal conduction and distribution of heat throughout the system. We stress that the Peltier and sample are much smaller in volume than the motors and the aluminum support. The latter two thus serve as a heat bath for our system with temperature equal to the ambient temperature.

## 3. Drift and system response

In conventional AFM, one of the main problems is the presence of thermal drift which leads to perceived movement of the lever that is imposed on the measured signal. Usually all parts in a system tend to move or expand as the temperature changes very slightly due to almost any source of heat that is in the vicinity. The transient response of drift may be as long as a day, being coupled to the heat influx from the sun. It may also be much shorter however as persons walk in or out of the room for example.

One way or the other, one has to deal with this drift effect and in many cases drift signals are easily identified by its slow change with time. For example in high precision force measurements electrical signals can be modulated at kHz frequencies, yielding a response that is many orders of magnitude faster than any thermal drift.

The above discussion is very relevant to our cause, for obvious reasons. Both heater and sample have dimensions in the order of $0.5cm^2$, yielding unavoidably slow modulation of heat. Smaller heaters are possible, but it severely degrades ease of use. Furthermore the heat of the Peltier is distributed throughout various parts of our system consisting of different materials with different heat capacity and conductivity. While the Peltier is very small in comparison with the rest of the system, and most parts are metallic ensuring optimal thermal conductivity throughout the system, yet the little amount of heat generated by operating the Peltier appeared to be enough to slightly change the long term drift that is normally present.

Fortunately we could make a distinction for the various signals related to different temperature sources with which we are dealing in our setup. Besides the laser whose thermal effect on the lever was not a problem as it was controlled by the feedback loop, we separate four kinds of mechanical motion connected to a thermal source.

(1) Motion of system parts due to external heat sources
(2) Motion of system parts induced by the measurement (heating the Peltier)
(3) Motion of the fiber/lever holder related to the radiative heat from the Peltier.
(4) Motion of the lever related to radiative heat from the Peltier/sample

The sources of motion in signal (1) and (2) could not be clearly identified, whereas we could correlate signal (3) and (4) to the thermocouples in our system. Ideally we wish to measure only signal (4). Signal (1) is present in any AFM setup to certain extends. It is shown in fig. 2a, and is typical in the order of 1nm per minute at room temperature. Signal (2) appears as soon as we start with the heating-cooling cycles (one cycle is 20 minutes), one can see that the long term drift is slightly modified after a few cycles (fig. 2b). Signal 2 is not easily identifiable with a single part of the system. It is also difficult to distinguish from signal (1). However we found it was always present, and correlated with the amount of heat that we put into our system. Whereas signal (1) could have constant direction for hours, signal (2) induced variations that became significant after about 2-10 minutes. Signal (2) is the reason that the presented measurements in fig. 2c,d loose correspondence with thermocouple signals over time.

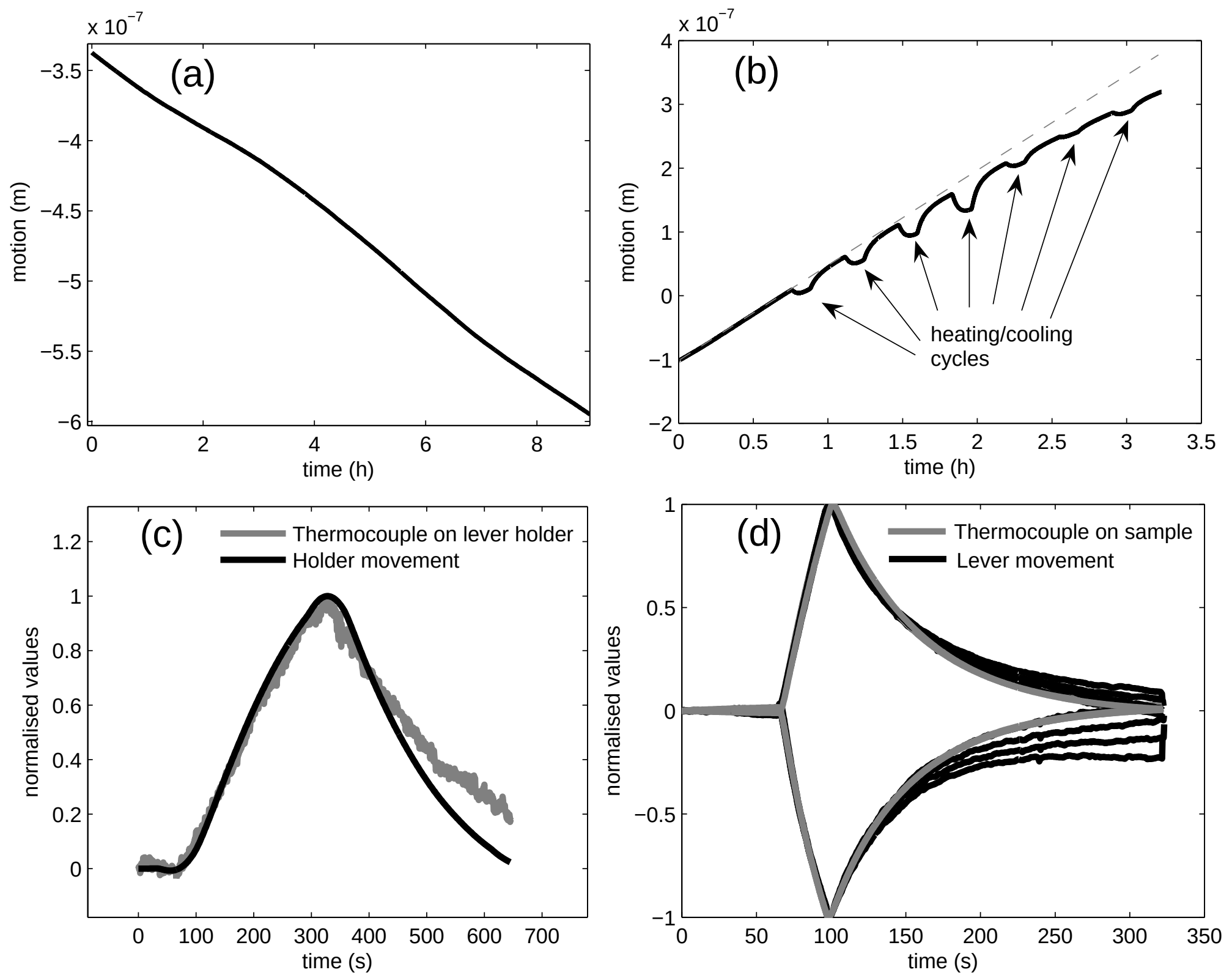

**Figure 2. (a)** Motion of long term drift over several hours. **(b)** The measurements (use of the Peltier) induce a change in the long term drift. **(c)** Due to radiated heat from the sample onto its vicinity, the aluminum lever holder expands. This can be detected by placing the fiber onto the holder. The temperature and expansion of the holder are correlated. **(d)** After correcting for drift, and subtracting the expansion signal (fig. 2c) we obtain lever motion that is well correlated with the temperature on the sample for both cooling and heating of the sample.

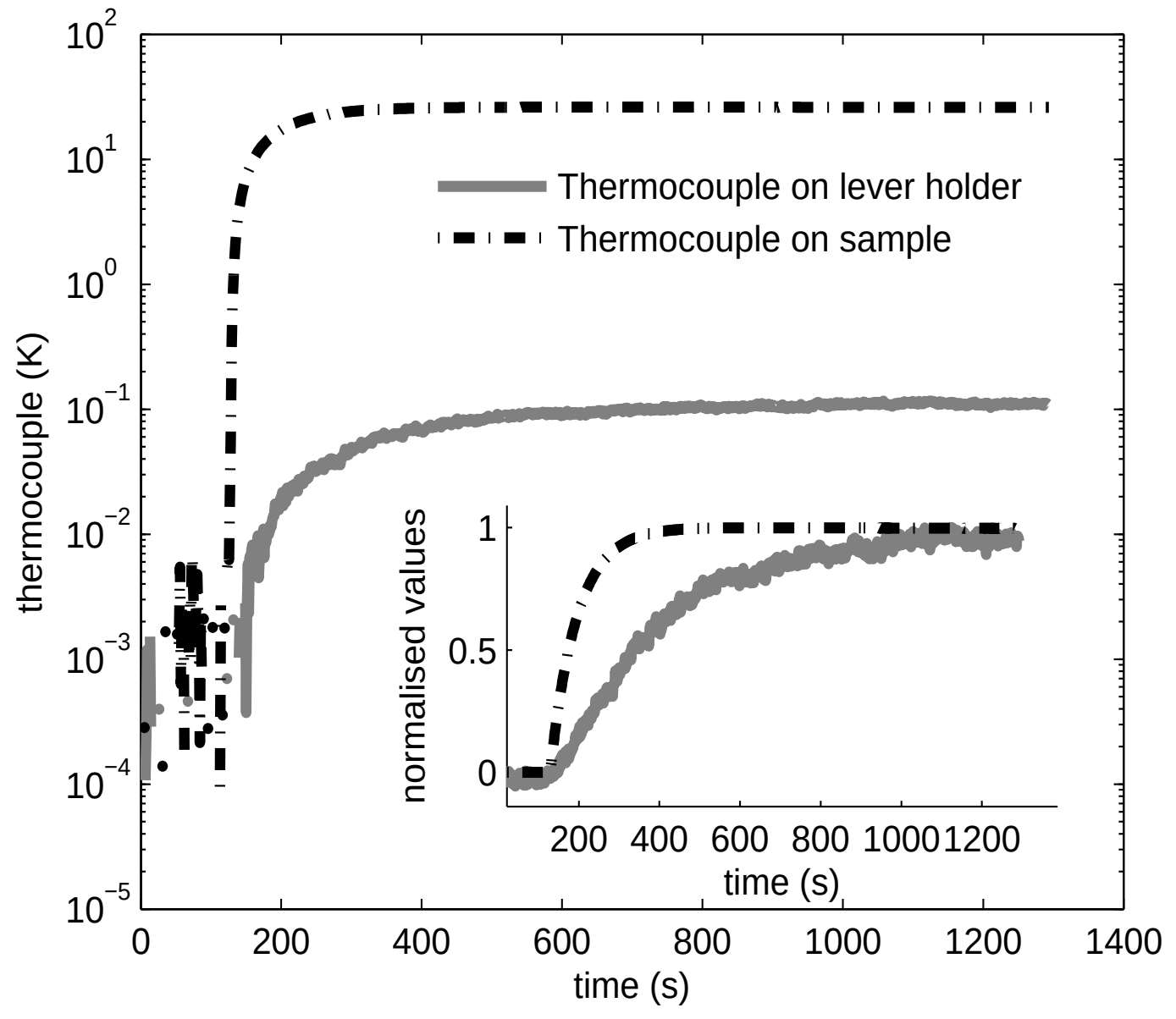

**Figure (3).** Measured temperature on the sample and on the lever holder while the sample is heated. The inset shows the normalized response to highlight the different transient responses.

The moveable fiber was not only used to measure lever deflection, but also to measure the movement of the lever holder. This yielded the measured signal (3) (fig. 2c, 3) which is related to heat of the sample/Peltier that is not only radiated unto the lever, but also unto the holder. We compared the movement of the lever holder to its temperature. When we heated the sample by 30K, the thermocouple on the lever holder registered a 0.12K temperature change that saturates after 20minutes (Figure 3). From figure 2c one can see that the movement of the lever holder, which in this case had a maximum of 60nm, and the signal of the thermocouple on the lever holder are correlated. Any differences in fig. 2c, that are visible after 400 seconds, are attributed to signal 2. Thus signal (3) is most likely related to thermal expansion of the lever holder. Using the formalism for thermal expansion of a material we obtain from $dL\cdot L^{-1}=\alpha dT$, with for aluminum $\alpha=23\cdot10^{-6}K^{-1}$, an estimate of the size of the holder of 1.6cm, which was very close to its actual size of about 2cm. We note here that also the fiber receives radiation from the sample. This signal depends on the position of the lever/fiber above the sample. At the edge of the sample the fiber will receive very little radiation. When placed in the middle however it receives a lot. We found our measurement to be reproducible regardless of the position of the lever above the sample. Thus both fiber and holder expansion are actually measured.

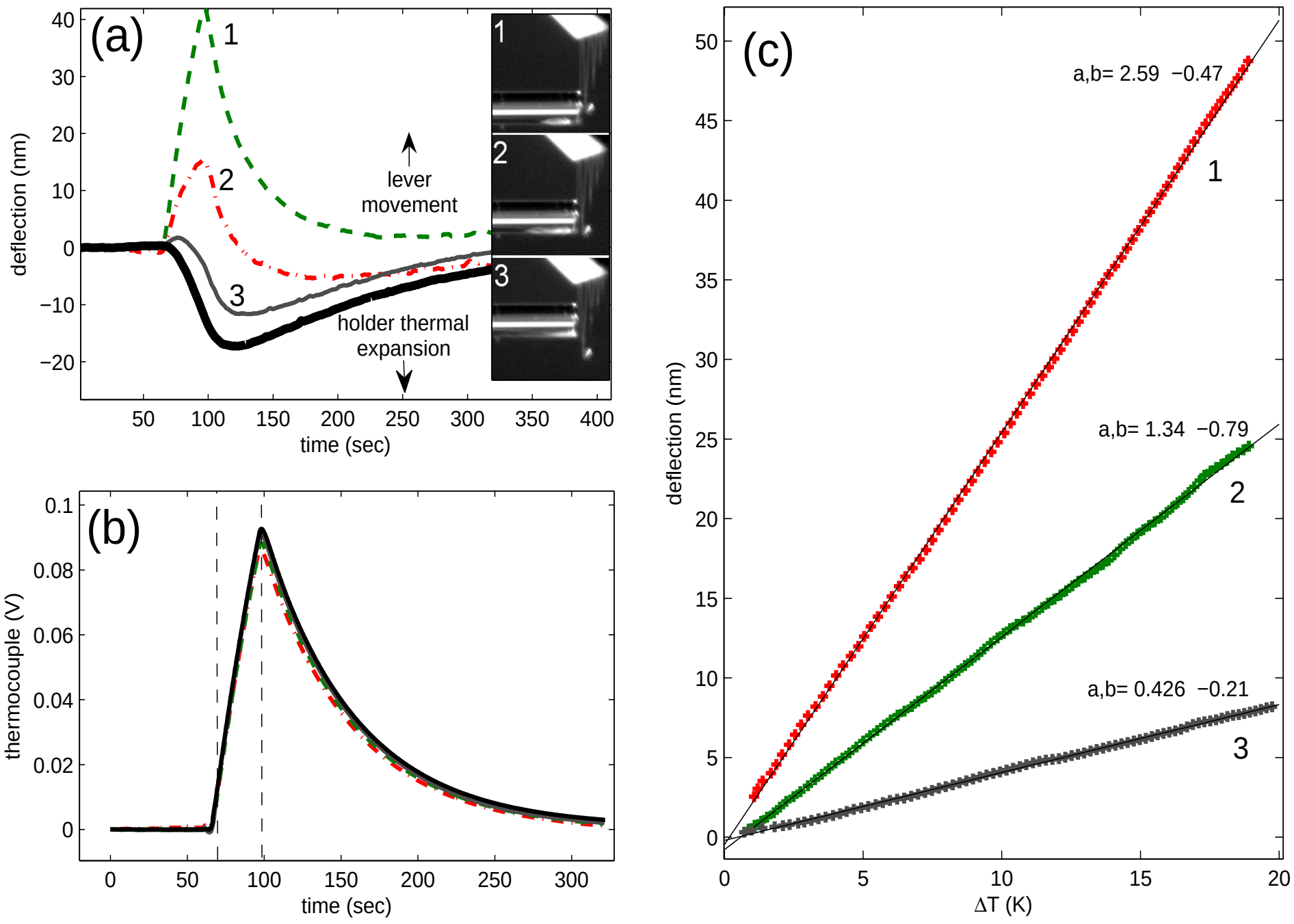

**Figure (4). (a)** The sensitivity of our setup depends on position of the fiber along the length of the lever. Positions 1-3 are shown in the inset. The fiber is moved vertically over the lever with a distance corresponding to the thickness of the fiber (125 micron). The lever holder movement is also shown. **(b)** Here the thermocouple response is shown, between the two dashed vertical lines one can determine the lever movement and relate it to the temperature difference between plate and sphere. **(c)** For small temperature differences as compared to the ambient temperature (300K) one can do a linear calibration with the function y=ax+b, where a is the sensitivity (nm/K) and b an offset. We found b to be zero within the measured variation when repeating the measurement.

To obtain signal (4) we put the fiber at the free end of the lever (fig. 4a inset1). Drift, or signal (1) was first fitted in the range 0-50 seconds. Then signal (3), which was measured under the same conditions but with the fiber on the lever holder, was subsequently subtracted from signal (1). One can see that there is a good correlation between the temperature on the sample and the lever motion, indicating that the lever moves due to radiative heat transfer. Due to signal (2), the correlation became worse after 150 seconds. We were not able to measure the transient response of the lever, as we could not change the temperature of the sample fast enough. Estimates from geometry of the sphere and lever however yielded transient response in the order of one millisecond [12]. Nonetheless we are now in a position to move on to a more quantitative analysis of our measurements.

## 4. Calibration

The calibration of the lever movement versus temperature is done as fast as possible to decrease the effect of the various drift mechanisms discussed in the last section. Typically it is done in 30 seconds as can be seen from figure 4b. The highest sensitivity to radiative heat transfer was obtained when the fiber measured the lever deflection at the free end of the lever (fig.4a inset 1). When the fiber was put 100μm higher or halfway on the lever (fig. 4a inset 3) the sensitivity had decreased six fold (fig. 4c). At the base of the lever we found practically no sensitivity. Once again, to obtain the curves in figure 4c, we first corrected for drift by fitting a linear function in the range 0-75 seconds. Then we subtracted the lever holder expansion signal (thick black curve in fig. 4a).

For a glass surface and 40μm glass sphere the holder expansion was about 3 times lower than the motion at the end of the lever, and had opposite sign. We varied the position of the fiber on the holder by about 1.5mm starting at the base of the lever, and found variations in the measured lever holder motion of 10% at most. Because the holder motion amounts to 30% of, and is subtracted from, the lever motion, it leads to a systematic error of 3% (10% of 30%) in the final motion of the lever. To assert the repeatability of our measurements we repeated it 8 times without changing lever or fiber position. For temperature differences that are not too large the lever motion versus applied temperature behaves linearly. By fitting the measured data with a function $y=ax+b$ as is done in figure 4c we can calibrate the lever sensitivity in nm/K. We obtained $a=1.51\pm0.015nm/K$, $b=-0.19\pm0.22nm$ for cooling, and $a=1.70\pm0.017nm/K$, $b=-0.07\pm0.5nm$ for heating in the range 0-20 Kelvin as obtained from 8 measurement runs performed the same place. Actual errors may increase when we include repositioning of the fiber on the lever. For samples with large emissivity such as glass and silicon we found a variation (standard deviation) in the obtained sensitivity at the 5-10% level from measurement to measurement and place to place when using this method. If repositioning is not needed the measured variations from curve to curve may be lower, and the systematic error dominates. For samples such as gold with very low emissivity the systematic error may be larger, also thermal noise becomes a larger problem.

The differences in the obtained sensitivities for cooling and heating are well explained by the Stefan-Boltzman law which for the radiated power P reads $P=A\varepsilon\sigma T^4$. Here A is the surface area, T the temperature, ε the emissivity of the material and σ the Stefan constant. In order to see the deviation from linearity at elevated temperatures better we heated the sample to about 110°C as is shown in fig. 5. This data can be fit with the function resembling the Stefan Boltzman law; $f(x)=-b\cdot c^4+b(\Delta T+c)^4$. Where b specifies a combination of cantilever sensitivity and material emissivity, and c is the ambient temperature (zero point in x-axis of fig. 5). The value found in this way for the ambient temperature was 410±30K, where 300K was expected. The deviation from linearity is still not very large in the measured temperature range. Furthermore for small ΔT (< 5K) the drift fitting procedure has some effect. When taking this into account by fitting $f(x)=a+b(\Delta T+300)^4$ we obtained good fits as well.

Farfield radiative heat transfer should not depend on the distance between the plate and the sphere. To test this we moved the sample using the Attocube Z-stage from about 50 micron to 3mm, and measured the farfield transfer to the lever. The results are plotted in fig. 6. Here the distance was estimated from optical images. A decrease of sensitivity with distance was observed. In our setup the lever with sphere is located about 1mm from the edge of the sample on one side, thus the observed effect is most likely due to a finite size (viewpoint) effect as the distance between the sphere and the plate is almost as large as the dimensions of the plate. In the distance range 50-200 micron the sensitivity changed by only 3% however.

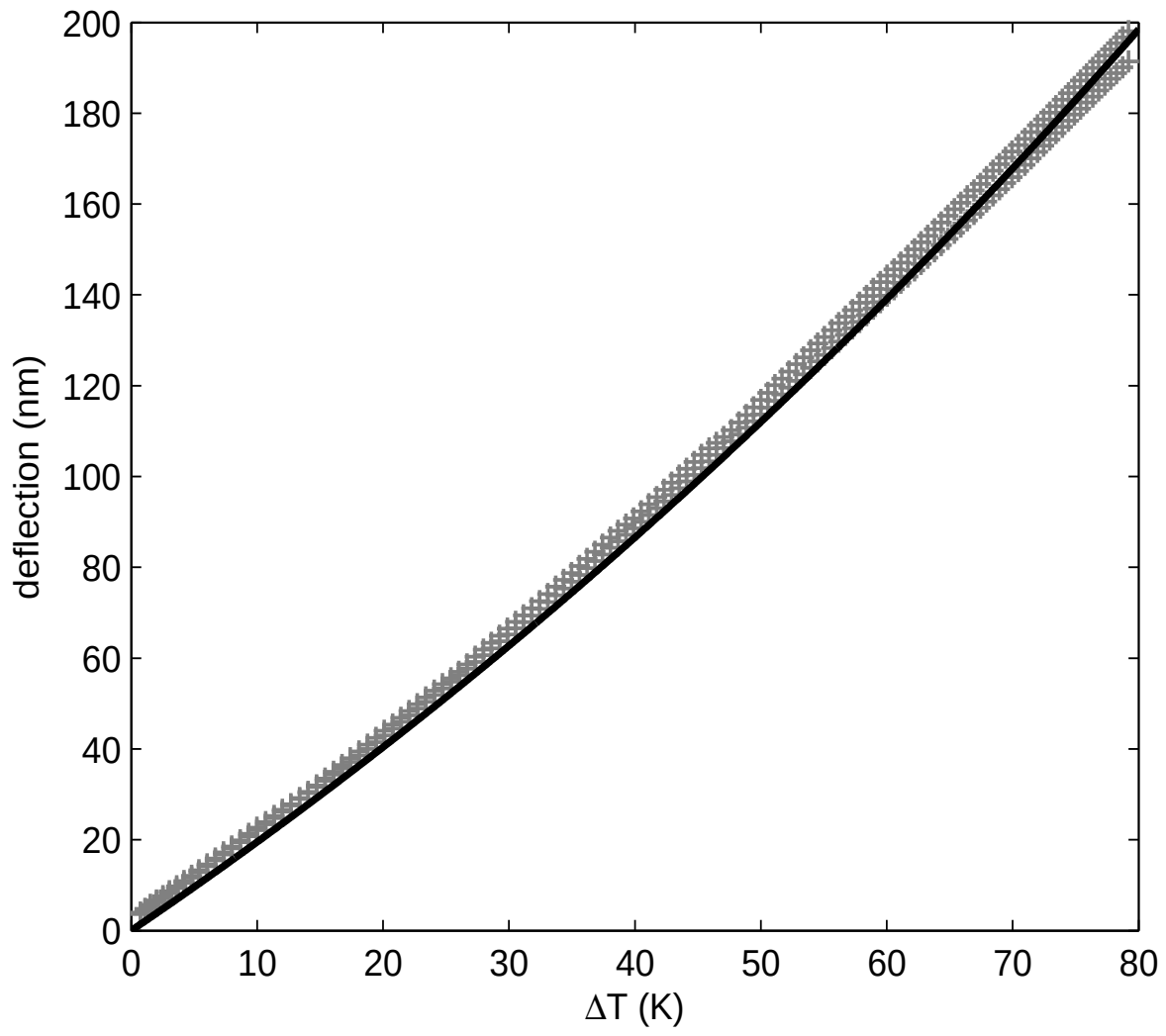

**Figure 5:** Measurements of lever deflection versus temperature difference between sphere and plate. The graph is similar to fig 4c, but the measurement is done over a larger range of ΔT. The black line is a fit of the Stefan-Boltzman law.

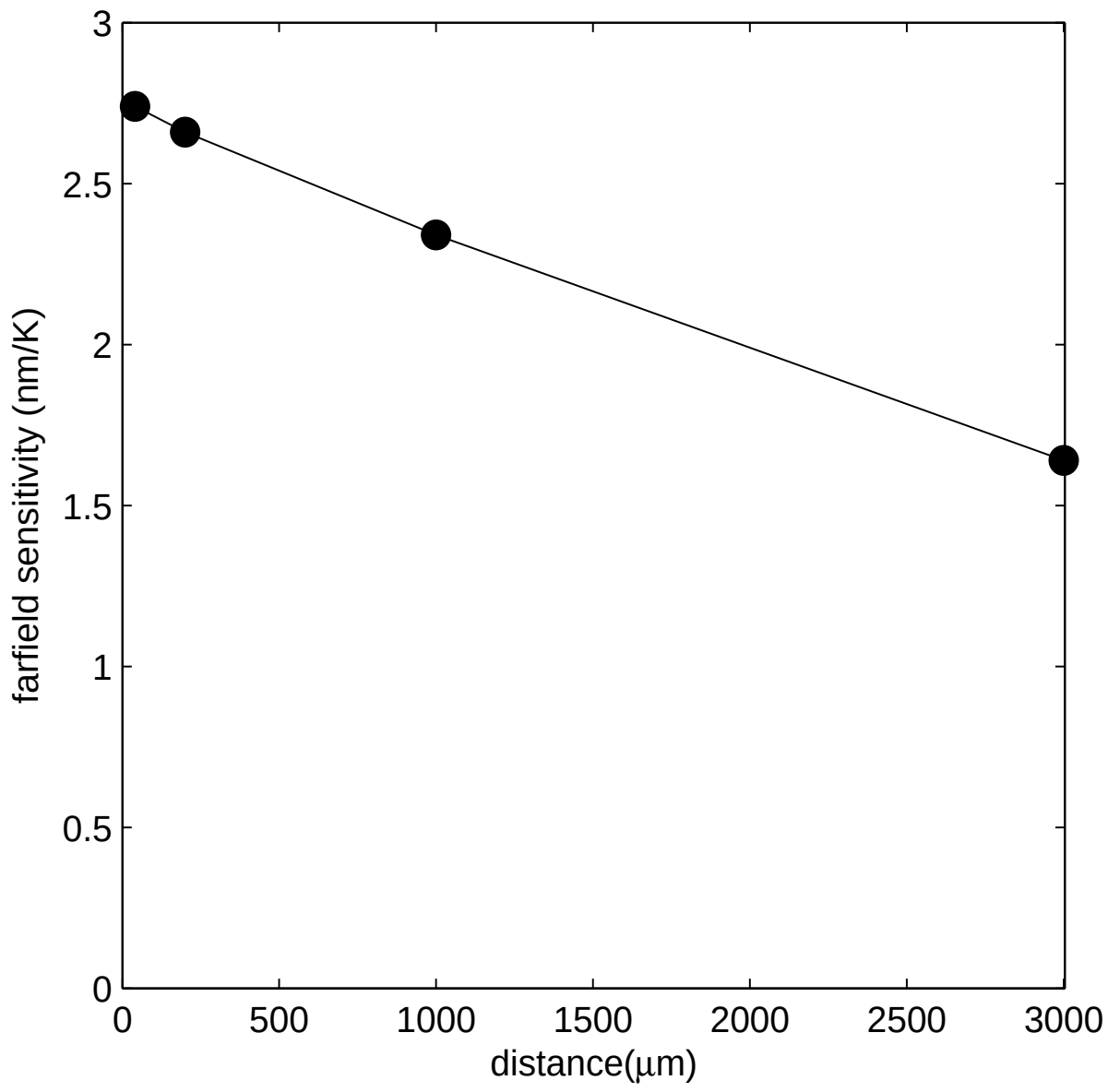

**Figure 6.** Farfield sensitivity of the lever with sphere versus the distance to the plate. The sensitivity decreases due to edge (viewpoint) effects as the distance between plate and sphere becomes similar in magnitude to the dimensions of the plate

## 5. Emissivity measurements

Farfield radiative heat transfer between two materials depends strongly on their dielectric properties. This material property is described by the emissivity ε. Following the Stefan Boltzman law the emissivity of a perfect blackbody is equal to one. For glass and silicon nitride it varies between 0.85-0.95, and for smooth Gold between 0.018-0.035. When heat is transferred between materials with different emissivity, the relation $\varepsilon_{1,2}= \varepsilon_1 \cdot \varepsilon_2/(\varepsilon_1+\varepsilon_2- \varepsilon_1 \cdot \varepsilon_2)$ holds. From this formula it follows that the least emissive materials dominate the final result. Thus the combined emissivity for gold-glass is almost equal to that of gold, being 24-50 times lower than that for glass-glass.

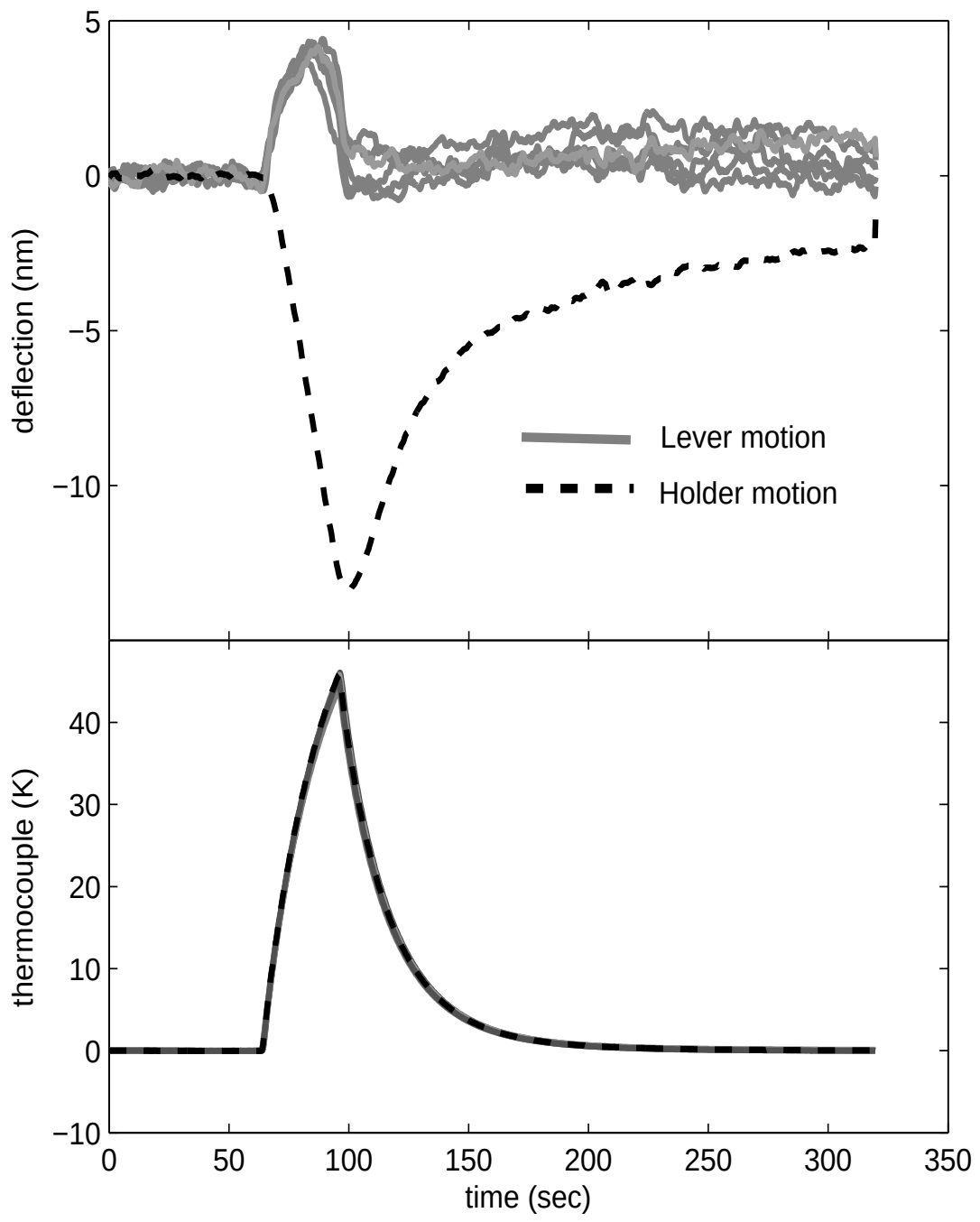

**Figure 7.** Motion of the lever (multiple curves) and the holder when using a gold plate instead of a glass plate. Note that both signals have significantly reduced as compared to using a glass plate. But the motion of the lever is now smaller than the motion of the holder. Note that we could heat more in this case as compared to figure 2, as the sample was much smaller.

Our results for a gold plate are shown in figure 7. The fiber was put at the free end of the lever (fig. 4a inset 1) to ensure maximum sensitivity. Both holder expansion and lever movement had significantly decreased. This is another sign that the lever holder expands due to thermal radiation from the heated sample. However the lever motion, being only 0.1nm/K, was found to be smaller than the expansion of the holder, and barely above the noise level. The detection limits of our setup were reached as the effects of drift fitting, and the subtraction of the holder motion now had strong effects. We measured that the farfield transfer decreased by a factor 25 in the case for glass-gold as compared to glass-glass (fig. 8). This is in good agreement with what is to be expected from the emissivity of the materials. We also performed measurements between silicon and glass (fig. 8). The measured heat transfer for silicon-silica was 30% lower as compared to silica-silica. As silicon has emissivity about 0.65, the combined emissivity is about 0.6, and is indeed about 30% lower than for silica-silica. Thus our apparatus is also sensitive to smaller emissivity differences between materials.

At this point the question may rise what the influence of the lever is on the farfield measurements, as not only the sphere but also the lever absorbs radiated heat from the sample. To test this we have performed a measurement between the same type of lever without sphere and a glass plate. A lever without sphere appeared indeed sensitive to heat transfer, but the sensitivity had decreased by a factor of 5 (fig. 8), indicating that the sphere dominated as the main heat absorbing part of the probe. Yet the sensitivity of a bare lever above a glass plate was found to be higher than that of a lever with sphere above a gold plate. This can be explained by the fact that the silicon nitride in the lever has high emissivity, which generates a significant effect in the case of a glass surface, but not in the case of gold.

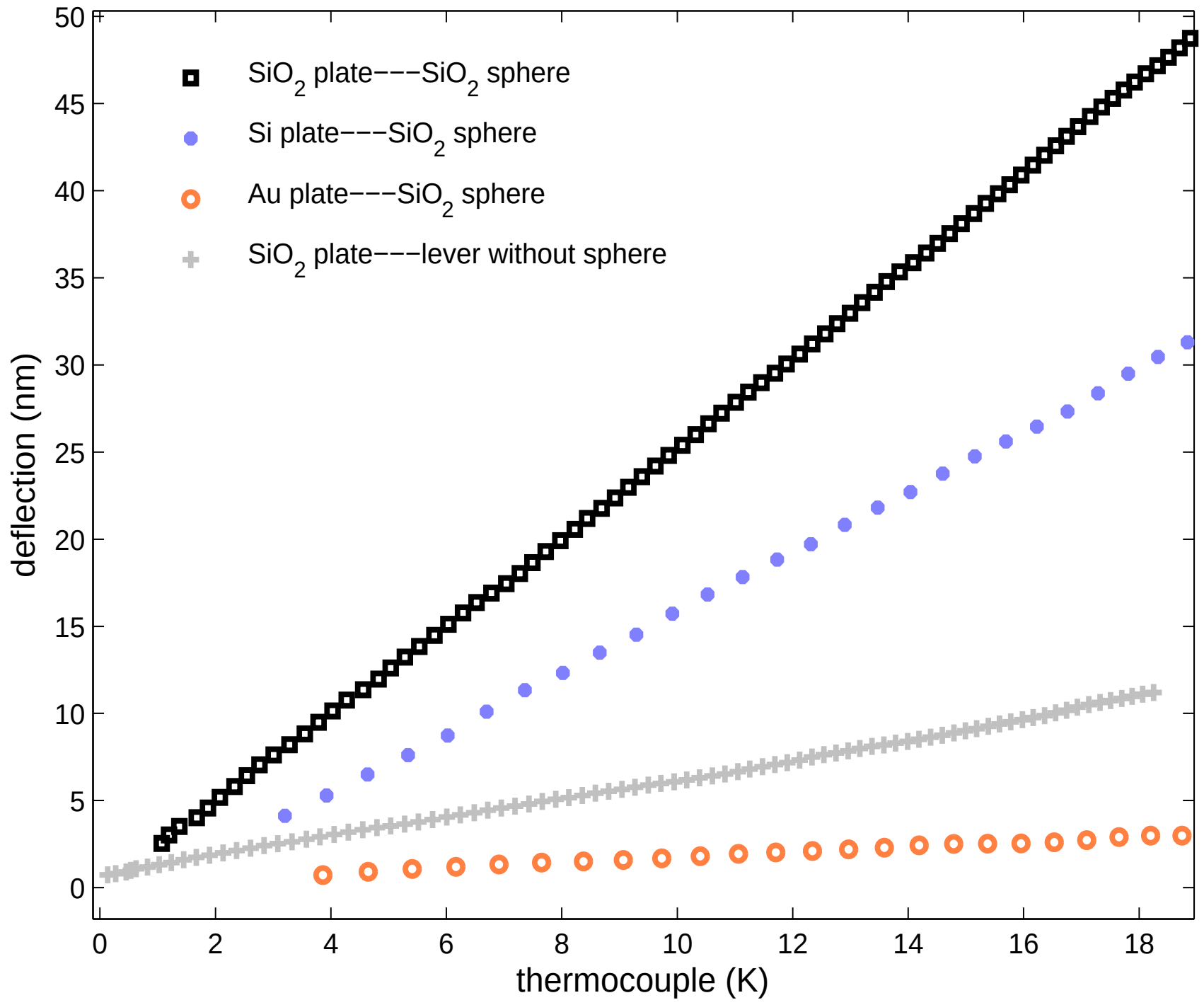

**Figure 8.** Lever deflection versus temperature difference between sphere and plate for different cases. This demonstrates the sensitivity of our setup for different materials. Note that a lever without sphere above a glass plate bends more due to applied heat than a lever with sphere above a gold plate.

## 6. Radiative heat transfer in farfield for $VO_2$

As an example of application we use our system to quantitatively probe the effect of a phase transition on radiative heat transfer between two objects. Vanadium dioxide undergoes a phase transition at 68°C, and exhibits large changes in optical and electrical properties. This renders it an important material for thermochromic windows [15], optical systems [16], and in electronic devices [17]. The emissivity of thin films of $VO_2$ changes by a factor 2 upon the phase transition [18], this should also affect the thermal transfer between two materials.

By means of laser ablation with a KrF excimer laser (20ns/pulse at 10Hz), a $VO_2$ thin film of approximately 100nm was deposited on sapphire. The sapphire was heated to 605°C with a heating rate of 20°C/min and deposition was done under $10^{-5}$mbar of Oxygen. Subsequently the film was cooled under 10mTorr oxygen atmosphere. The reflectivity of this $VO_2$ film increased by 35% whereas its conductivity changed by almost 3 orders of magnitude upon the phase transition. The roughness of the film was measured by AFM to be 12nm rms over a 4 micron$^2$ area.

We measured heat transfer between the $VO_2$ film and a glass sphere mounted on the lever as was done before. The transition temperature was clearly visible in the form of a peak in the lever motion curve when heating with $\Delta T$=44K (fig. 9). As the ambient temperature was 297K we measured the phase transition temperature of this film to be 341K, which is the expected value [19]. The effect of the phase transition was also visible on the lever holder, but it was much less pronounced (fig. 9).

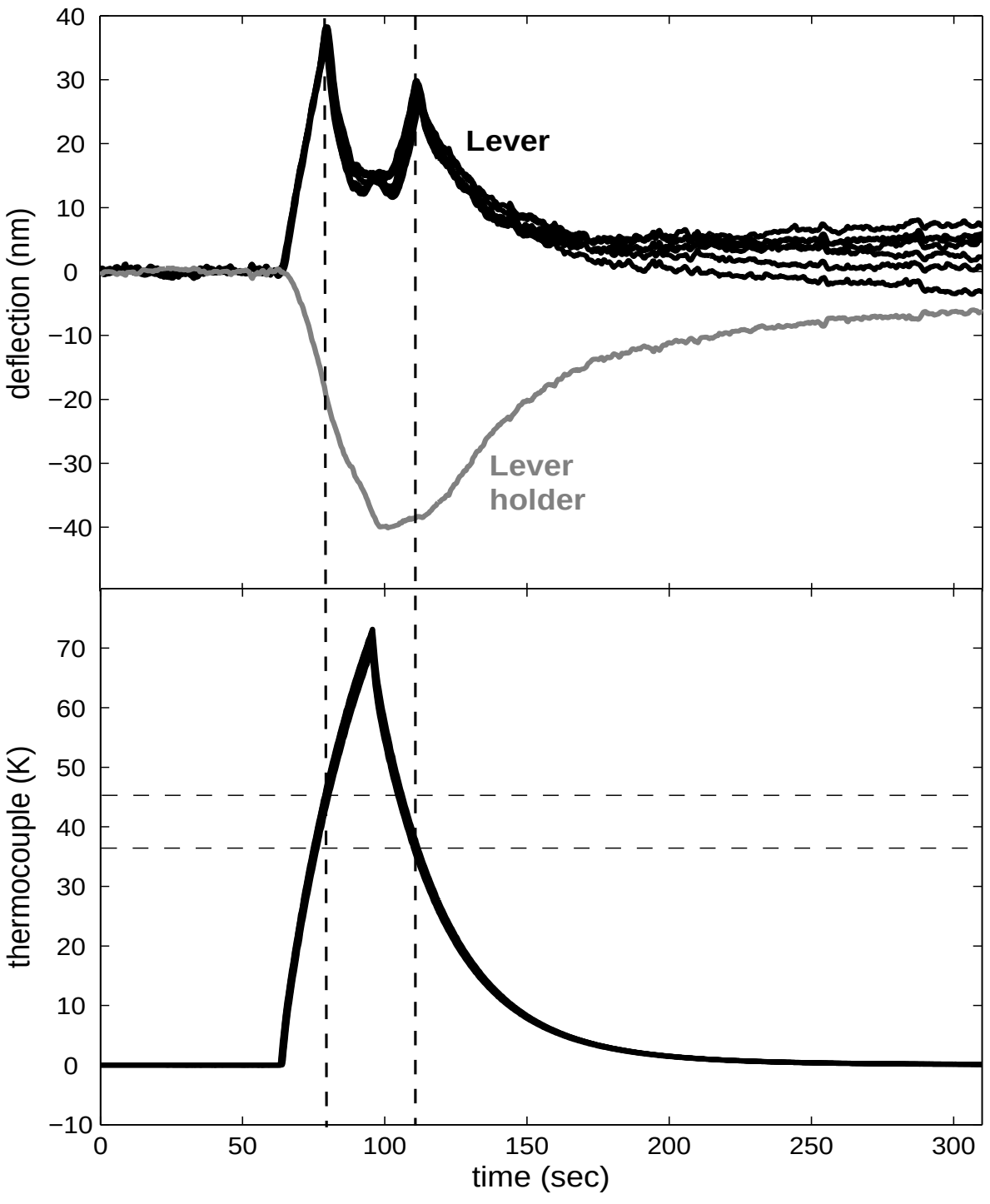

**Figure 9.** Multiple measurements of heat transfer between a lever with 40 micron glass sphere and a $VO_2$ sample that undergoes a phase transition. A clear hysteresis effect is visible for the transition temperature. Once again the effect of drift becomes visible after 150 seconds.

Below the transition temperature we observed an almost linear increase of flux with temperature. Thus the thermal transfer divided by the temperature remained almost constant under the transition temperature (fig. 10 inset). A slight increase was seen from about 5K below the transition temperature (fig. 10 inset). This is intriguing as such effects are not seen for either conductivity or reflectivity measurements. At the phase transition temperature the thermal transfer drops sharply and continues decreasing by as much as a factor of 5 at 370K, most likely due to $VO_2$ becoming more conductive and reflective. Furthermore a clear hysteresis effect is visible in fig. 10, as the transition temperature is 10K lower when cooling. This is in good agreement from what is known from conductivity and optical measurements [18, 19]. Note that the behavior observed here corresponds closely with the measured behavior of the emissivity of $VO_2$, however it is more pronounced. Whereas the emissivity changed only by a factor of 2 in ref. [18], we measured a change in heat transfer of a factor of 5. This highlights the importance of interaction materials, and serves as a unique example of application of the technique presented in this work.

To explain the magnitude of the observed effects we performed farfield radiative heat transfer calculations for $VO_2$-glass and $VO_2$-gold systems, by using dielectric data for gold, glass and vanadium dioxide and the methods in refs. [20, 21]. For $VO_2$-glass we found a change of a factor of 5.5 upon phase transition, which is in good agreement with what we measured. For $VO_2$-gold the calculated effect was very small as the thermal transfer changed by less than 5% upon the phase transition. This explains the small effect observed for the holder, which is also a metal. Although we feel we cannot make quantitative comparisons for this case.

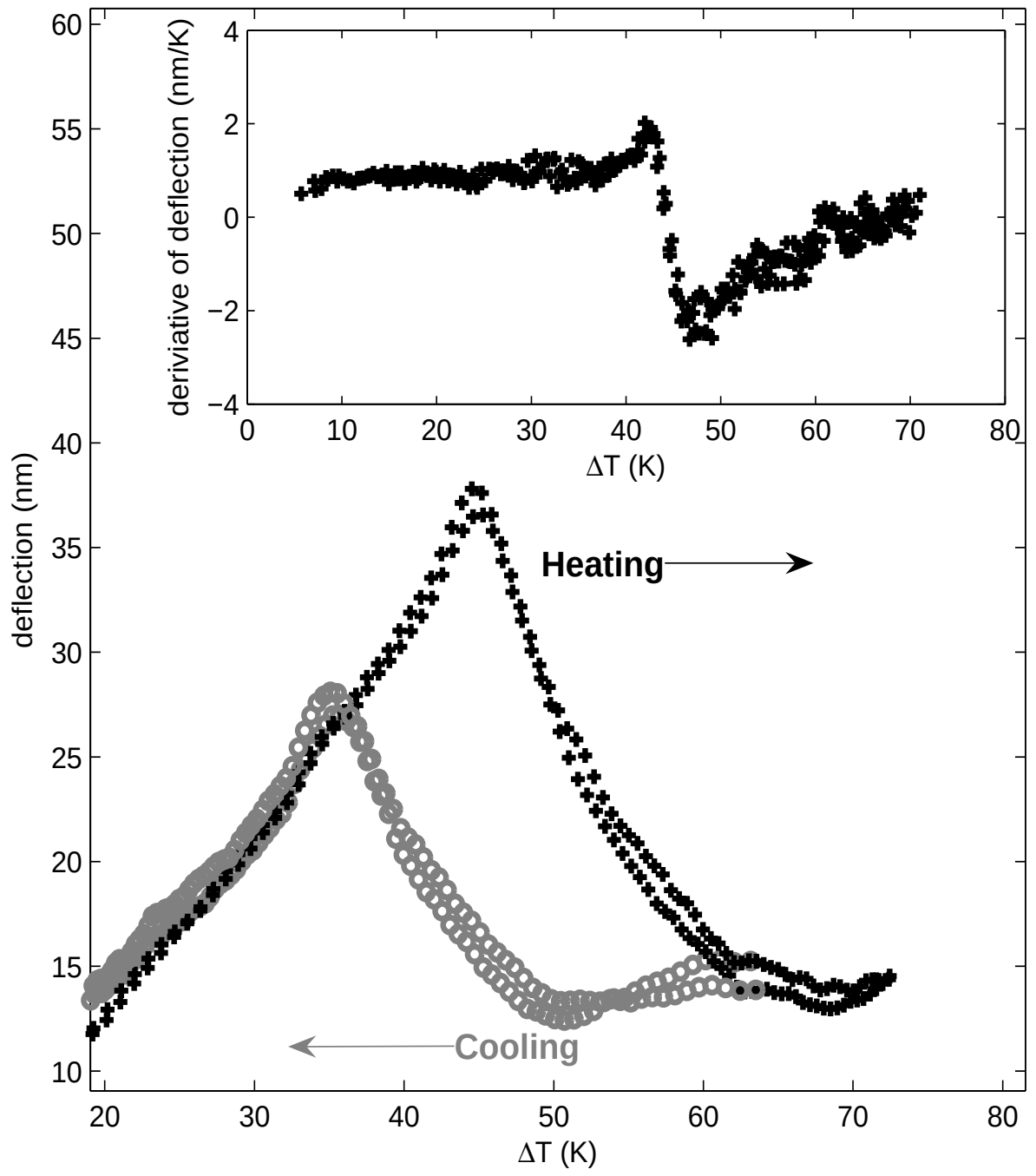

**Figure 10.** The lever motion is plotted versus the applied temperature difference between the sphere and the sample. The heat transfer decreases by a factor of 5 across the phase transition. The hysteresis effect is clearly visible. Drift affects the measurement with time, which becomes more visible for the cooling part of the data. The inset shows the derivative Δdeflection/Δtemperature.

## 7. Conclusion and outlook

We have given extensive details of an AFM that is very well suited for the measurement of radiative heat transfer due to the inclusion of extensive measures that reduce the heating effects of the optical detection system. This system has a large ease of use in near and far-field, as compared to measurements between parallel plates [22,23]. We demonstrated it by performing heat transfer curves versus temperature and found good correspondence with the Stefan Boltzman law. The lever with glass probe can be calibrated to a material with known emissivity to yield quantitative radiative heat transfer, temperature or emissivity information of other samples. We have demonstrated this ability by measuring farfield thermal transfer for different materials, such as gold and glass and probed the effect of the phase transition of $VO_2$ on radiative transfer in farfield to a glass probe, which changed reversibly by a factor of 5.

The setup in its present state offers the same ease of use as a conventional AFM, however as the sample is heated, radiative heat also warms up the surroundings. We found this heat effect to be very small being 0.3% of the temperature applied to the sample. Yet this slight temperature increase leads to thermal expansion of the lever holder, and most likely other parts of the system. While we have shown that this can be well characterized, the performance of the system can be improved by the use of FeNi36 [24] Invar alloys that have a thermal expansion coefficient of $0.6 \cdot 10^{-6} K^{-1}$, which is a factor 35 lower than that of aluminum. In this case the thermal expansion signal would comprise less than 1% of the measured signal. Furthermore the sensitivity of bilayer levers can be much improved, for example by depositing polymer layers on commercially available polysilicon levers [10]. For gold-silicon nitride levers the difference in thermal expansion coefficients is about $10^{-5} K^{-1}$, for polymer-polysilicon levers this is increased by a factor of 30 [10]. Foremost this yields an increased dynamic range so that it can be used to measure more accurately the relatively weak thermal transfer for metal-metal systems. On the other hand one can also choose to use correspondingly smaller probes that allow high density high sensitivity MEMS arrays of micro levers that may one day replace expensive and bulky cryogenic IR detectors [11]. Functionalized probes at the end of such levers, as well as a stabilized optical detection system such as the one presented here, may greatly enhance the accuracy and sensitivity of such devices.

In summary we have presented a technique based on bilayer levers that is able to accurately probe near and far field radiative heat transfer. We have used this to determine how the phase transition of $VO_2$ changes radiative heat flux to a glass probe. Work is in progress to probe the phase transition of $VO_2$ in near field with this technique.

**Acknowledgements**
We gratefully acknowledge support of the Agence Nationale de la Recherche through the Source-TPV project ANR 2010 BLAN 0928 01. We acknowledge fruitful discussions with K. Joulain.

## References

[1] J.Currano,S. Moghaddam, J. Lawler, K. Jungho, J. Thermo. Heat Transfer, 22 360 (2008)

[2] R. J. Pylkki, P. J. Moyer, P. E. West, Jpn. J. Appl. Phys. 33 3785 (1994)

[3] J. Christofferson, K. Maize, Y. Ezzahri, J. Shabani, X. Wang, and A. Shakouri, J. Electron. Packag. **130**, 041101 (2008).

[4] Rogalski, *Infrared Phys. Technol. 35*, 1-21 (1994).

[5] C. Williams and H. K. Wickramasinghe, Appl. Phys. Lett 49 1387 (1986)

[6] Kittel, W. Müller-Hirsch, J. Parisi, S. A. Biehs, D. Reddig, M. Holthaus, Phys. Rev. Lett. 95, 224301 (2005).

[7] Y. De Wilde, F. Formanek, R. Carminati, B. Gralak, P.-A. Lemoine, K. Joulain, J.-P. Mulet, Y. Chen, and J.-J. Greffet, Nature (London) **444**, 740 (2006).

[8] J. R. Barnes, R. J. Stephenson, C. N. Woodburn, S. J. O'Shea, M. E. Welland, T. Rayment, J. K. Gimzewski, and Ch. Gerber, Rev. Sci. Instrum. **65**, 3793 (1994).

[9] O. Nakabeppu, M. Chandrachood, Y. Wu, J. Lai, and A. Majumda, Appl. Phys. Lett. 66, 694 (1995).

[10] M. C. LeMieux, M. E. McConney, Y.-H. Lin, S. Singamaneni, H. Jiang, T. J. Bunning, and V. V. Tsukruk, Nano Lett. **6**, 730 (2006).

[11] L. R. Senesac, J. L. Corbeil, N. V. Lavrik, S. Rajic, and P. G. Datskos, Ultramicroscopy **97**, 451 (2003).

[12] P. G. Datskos, N. V. Lavrik, and S. Rajic, Rev. Sci. Instrum. **75**, 1134 (2004)

[13] Narayanaswamy, S. Shen, and G. Chen, Phys. Rev. B **78**, 115303 (2008).

[14] E. Rousseau, A. Siria, G. Jourdan, S. Volz, F. Comin, J. Chevrier, and J.-J. Greffet, Nat. Photonics **3**, 514 (2009).

[15] N. R. Mlyuka, G. A. Niklasson, and C. G. Granqvist, Appl. Phys. Lett. **95**, 171909 (2009).

[16] M. Rini et al, Opt. Lett. **30**, 558 (2005)

[17] A.Crunteanu et al, Sci. Technol. Adv. Mater. 11 (2010) 065002

[18] F. Guinneton, L. Sauques, J.C. Valmalette, F. Cros, and J.R. Gavarri, J. Phys. Chem. Solids **62**, 1229 (2001).

[19] E. E. Chain, Appl. Opt. **30**, 2782 (1991)

[20] P. J. van Zwol, K. Joulain, P. Ben-Abdallah, J. -J. Greffet, and J. Chevrier, Phys. Rev. B 83, 201404(R) (2011)

[21] P.J. van Zwol, K. Joulain, P. Ben-Abdallah, J. Chevrier, Phys. Rev. B 84, 161413(R) (2011)

[22] R. S. Ottens, V. Quetschke, Stacy Wise, A. A. Alemi, R. Lundock, G. Mueller, D. H. Reitze, D. B. Tanner, and B. F. Whiting, Phys. Rev. Lett. 107, 014301 (2011)

[23] T. Kralik, P. Hanzelka, V. Musilova, A. Srnka, and M. Zobac, Rev. Sci. Instrum. 82, 055106 (2011)

[24] M. Uhl, L. M. Sandratskii, and J. Kübler, Phys. Rev. B **50**, 291 (1994).